\documentclass[a4paper,11pt]{article}
\pdfoutput=1 

\usepackage{jinstpub} 

\usepackage{lineno}

\usepackage{graphicx}

\usepackage[dvipsnames]{xcolor}
\usepackage{lineno}

\usepackage{amsmath,amstext,amsthm,amsfonts,amsfonts}
\usepackage[utf8]{inputenc}
\usepackage{subcaption}
\usepackage[linesnumbered,ruled]{algorithm2e}
\usepackage{url}
\usepackage{soul}
\usepackage{bm}
\usepackage[normalem]{ulem}
\usepackage{multirow, makecell}

\usepackage{hyperref}

\usepackage{amsmath}

\newcommand{\sonoarrivatoqui}[1]{\textcolor{red}{\\ \%\%\% sono arrivato qui \%\%\% \\\\}}

\DeclareMathAlphabet{\mathsfit}{\encodingdefault}{\sfdefault}{m}{sl}
\SetMathAlphabet{\mathsfit}{bold}{\encodingdefault}{\sfdefault}{bx}{n}

\makeatletter
\def\@fnsymbol#1{\ifcase#1\or *\or i\or ii\or iii\or iv\fi}
\makeatother

\begin{document}

\title{A 3D segmented Water-based Liquid Scintillator for high-precision detection of neutrinos in water}

\author[a]{Botao Li,} 
\author[a, ii]{Daria Borodulina,\note{Now at Laboratoire De Physique des deux infinis (LP2i), CNRS/IN2P3, Boardeaux, France}}
\author[a, i]{Davide Sgalaberna,}
\author[a]{Tim Weber,}
\author[b]{Minfang Yeh,}
\author[a]{Umut Kose,}
\author[a]{Matthew Franks,}
\author[a]{Andr\'e Rubbia,}
\author[a, iii]{Johannes Wüthrich,\note{Now at the University of Zurich, Switzerland}}
\author[d]{Claudio Giganti,}
\author[c]{Tatsuya Kikawa}

\note{Corresponding author.}

\affiliation[a]{ETH Zurich, Institute for Particle Physics and Astrophysics, CH-8093, Zurich, Switzerland}
\affiliation[b]{Chemistry Division, Brookhaven National Laboratory, Upton, NY 11973, USA}
\affiliation[c]{Kyoto University, Department of Physics, Kyoto, Japan}
\affiliation[d]{Laboratoire de Physique Nucleaire et des Hautes Energies (LPNHE), Sorbonne Universit\'e, Universit\'e Paris Diderot, CNRS/IN2P3, Paris, France}

\emailAdd{davide.sgalaberna@cern.ch}

\abstract{
Precision detection of neutrino-nucleus interactions in water with the complete detection of the final state, including leptons and hadrons, is challenging due to water being a non-scintillating medium. 
This can be a limitation for the next-generation long-baseline neutrino oscillation experiments,
such as Hyper-Kamiokande, 
where the neutrino-nucleus interaction models must reach a few percent-level accuracy.
Water-based liquid scintillator can be a game changer for the future near detectors.
In this article, we propose a novel design consisting of a 3D highly-segmented water-based liquid scintillator. The water-based liquid scintillator is encapsulated within a highly-segmented rigid but very light structure that provides the optical isolation with a 1 cm$^{3}$ granularity, each read out by orthogonal wavelength shifting fibers, and 81\% of water by mass in the active volume.
Such configuration is also suitable for pure liquid scintillator. 
The detector design, prototyped and validated with cosmic ray data, is described and results are reported. The optical model is studied with Monte Carlo simulations and results are compared with the collected data.
}

\maketitle

\section{Introduction}
\label{sec:introduction}

\label{sec:motivation}
Water-based detectors are widely used as a neutrino active target 
due to the low cost and the scalability to very large sizes
~\cite{SNO:2002tuh,Kamiokande:1994sgx,t2k,Hyper-Kamiokande:2018ofw}.
On the other hand, only Cherenkov radiation is produced in water and, thus, particle tracking remains difficult, especially for nucleons, often below the detection threshold at the energies typical for current neutrino experiments. Moreover, the intensity of the Cherenkov radiation is about two orders of magnitude lower than typical scintillation yields in organic scintillators \cite{Kolanoski:2020ksk}. 
In current and future long-baseline neutrino oscillation (LBL) experiments, such as T2K \cite{t2k} and Hyper-Kamiokande \cite{Hyper-Kamiokande:2018ofw,hyperk-input-esppu}, a multi-kiloton water-Cherenkov detector serves as a far detector (FD) to measure the neutrino event rate at the maximum of the flavor oscillation probability. 
For these experiments, the ability to accurately compare the neutrino event spectrum measured at the FD with that obtained at a near detector (ND), before oscillations can occur, is critical for reducing systematic uncertainties related to the modeling of the neutrino flux and the interaction cross section, and for achieving precise oscillation measurements.
At Hyper-Kamiokande, whose goal is to measure the leptonic charge-parity violating phase ($\delta_{CP}$), the requirement of a percent-level cross-section uncertainty highlights the need of a water-based ND \cite{hyperk-sensi}
to constrain possible systematic uncertainties arising from differences in the nuclear properties of oxygen and carbon or other constituent nuclei affecting the final state of neutrino interactions \cite{Abe:2020uub,MINERvA:2023kuz}.
%
The precision detection of neutrino-nucleus interactions in water with the complete detection of the final state, including leptons and hadrons, has always been a challenge given that water does not scintillate. 
While an obvious solution 
to improve the sensitivity 
is to deploy a large-volume water-Cherenkov ND \cite{Hyper-Kamiokande:2018ofw,hyperk-sensi}, this configuration is not sensitive to all the final-state particles and, thus, misses information important for achieving a complete characterization of neutrino interactions, otherwise available in standard scintillator-based detectors.

The ideal configuration would be given by a highly-segmented scintillating detector made of water and capable of tracking, identifying, and measuring the stopping power of various types of ionizing particles with the typical energy threshold of organic scintillators.
The solution  
is provided by water-based liquid scintillator (WbLS) \cite{WbLS}, 
that integrates organic scintillators into a water matrix. It combines a high fraction of water with the properties of liquid scintillator, required by a ND for the detection of hadrons like protons below 1 GeV/c. 
Its tunable concentration allows for optimization to meet specific experimental requirements, making it a versatile detection medium. The feasibility of the WbLS for ton-scale applications has been recently demonstrated \cite{ton_scale_wbls1, ton_scale_wbls2, ton_scale_wbls3}.

In this work we propose a novel design that consists of a 3D fine granularity WbLS detector, whose geometry is inspired by the plastic scintillator detector made of about 2,000,000 cubes \cite{Sgalaberna:2017khy}, recently installed in Japan to upgrade the T2K magnetized ND (ND280) \cite{ND280upgrade-tdr}.
Despite the fine segmentation, in the proposed WbLS detector a high fraction of water is achieved, thus allowing to minimize the systematic uncertainty that results from the ``subtraction'' of contributions of other nuclei, such as carbon, in the measurement of the neutrino-water cross-section.
The design and the performance characterization of a small-scale 3D segmented WbLS 
prototype is reported in this work. Emphasis is given to  scintillation properties such as light yield in response to cosmic ray events and the voxel-to-voxel optical crosstalk. Comparative measurements were also performed using a standard 
liquid scintillator with the same composition as that 
designed for the Daya Bay experiment \cite{DayaBay}.

This novel configuration can be instrumental for the Hyper-Kamiokande ND to complement 
its 
intermediate water-Cherenkov detector (IWCD) 
\cite{hyperk-sensi,Abe:2018uyc}
and the magnetized near detector, ND280, whose neutrino target material is mostly plastic scintillator.
For instance, Hyper-Kamiokande is considering the possibility of an additional future upgrade of ND280 for its high-statistics phase
 \cite{hyperk-input-esppu}.
Similarly, a 3D segmented WbLS detector could be adopted for the proposed LBL ESS$\nu$SB experiment \cite{Alekou:2022emd}
or 
serve as ND 
for THEIA, a 25 kiloton WbLS detector proposed as the fourth module at the DUNE LBL experiment 
\cite{DUNE:2024wvj}.

\section{Design of a 3D-segmented water-based liquid scintillator with large water content}
\label{sec:conceptual-design}

The novel design has to achieve the following goals: 3D optical isolation of WbLS with 1~cm granularity (range of protons in water with Fermi momentum of oxygen), 
a low-energy threshold sufficient
to detect minimum ionizing particles (MIP) and protons, a water fraction by mass > 80\% and both light yield and mechanical stability. 
The WbLS was produced at Brookhaven National Laboratory (BNL).
The chosen composition consists of 90\% of water by mass loaded with 10\% of the active component, that is 
linear alkyle-benzene (LAB) with 3 g/L PPO (2,5- diphenyloxazole) and 15 mg/L MSB (1,4- Bis (2-methylstyryl) benzene).
It provides a nominal light output of about 9,000 to 12,000 photons / MeV,
depending on the vendor and the method of purification,
with a decay time of about 2 ns \cite{DayaBay,SNO+,Theia:2019non}. 
The light yield of WbLS is approximately proportional to the fraction of organic scintillator in water~\cite{WbLS,Theia:2019non}, thus it is expected to be around 900-1,200 photons/MeV.
The absorption length in WbLS as well as in pure liquid scintillator LAB+PPO+MSB (LS) is of the order of a few tenths of meter~\cite{WbLS, LS}, thus the attenuation of light in a 1 cm$^3$ voxel is negligible. 
The optical separator must provide mechanical rigidity to the structure 
but shall introduce a minimal amount of non-water material.
Thus, it is made of a core of low-density Divinycell H80~\cite{foam} foam (0.08 g/cm$^3$) with a thickness of 1.2 mm and a height of 10 mm. It is sandwiched between two 3M DF2000MA specular multilayer polyester-based adhesive films, each one about 100 $\mu$m thick, that reflects in air $>$99\% of visible light at normal incidence ~\cite{3m_df2000ma}. 
The 3M specular films are widely used in liquid, plastic and inorganic scintillator detector~\cite{prospect, Plas_Det_3M, CsI_Det_3M}.
The total thickness of the optical separator is 1.4 mm.
Slits (1~mm in width and 5~mm in height) were made with a pitch of 10~mm on the vertical separators and allowed to connect them into a rigid layer consisting of a mechanically-constrained horizontal grid of 1~cm$^2$ voxels. Two adjacent layers, one on top of the other, were sub-divided by a horizontal separator plane, resulting in a very efficient optical isolation that forms the 3D WbLS voxels while keeping the overall structure at low density but rigid.
Holes in the optical separators were made to host the wavelength shifting (WLS) Kuraray Y11 double-cladding fibers of 1~mm diameter along two orthogonal directions, that 
transit 
the scintillation photons produced in the WbLS 
to the silicon photomultipliers (SiPM), placed outside of the box. 
1.05~mm diameter holes were sealed by the use of rubber O-rings compressed by aluminum plates screwed on the box walls. A cap closed the top side of the wall. 
All the detector optical components (separators and WLS fibers) were immersed in the WbLS inside the water-tight dark acrylic box.

The result is a 3D matrix of WbLS. 
Each cube has an inner dimension of 1 $\times$ 1 $\times$ 1 $cm^3$ and is filled with WbLS. Two cubes are separated by a 1.4 mm thick optical separator, with an average density per cube of about 0.73 g/cm$^3$.
If we exclude the water-tight dark box, the water mass ratio reaches 81\% thanks to the very low density of the reflector frame. Moreover, all the materials immersed in the WbLS (3M mirror, Divinycell, WLS fibers) are organic materials, thus predominantly composed of carbon, whose neutrino-nucleus cross section is being extensively studied in literature \cite{MINERvA:2023kuz,MINERvA:2022mnw,ND280upgrade-tdr,T2K:2023qjb}. 
The organic part is dominated by the fraction of LS loaded into the WbLS (10\%). Only a small fraction is taken by the WLS fibers (about 1.5\% for a total of two fibers) and the optical separators (about 3.5\%).  
The small quantity of carbon allows to subtract its contribution to the number of neutrino interactions with a minimal impact from propagation of the systematic uncertainties related to the nuclear models.

\begin{figure}[h!]
	\centering 
    \includegraphics[width=0.9\textwidth]{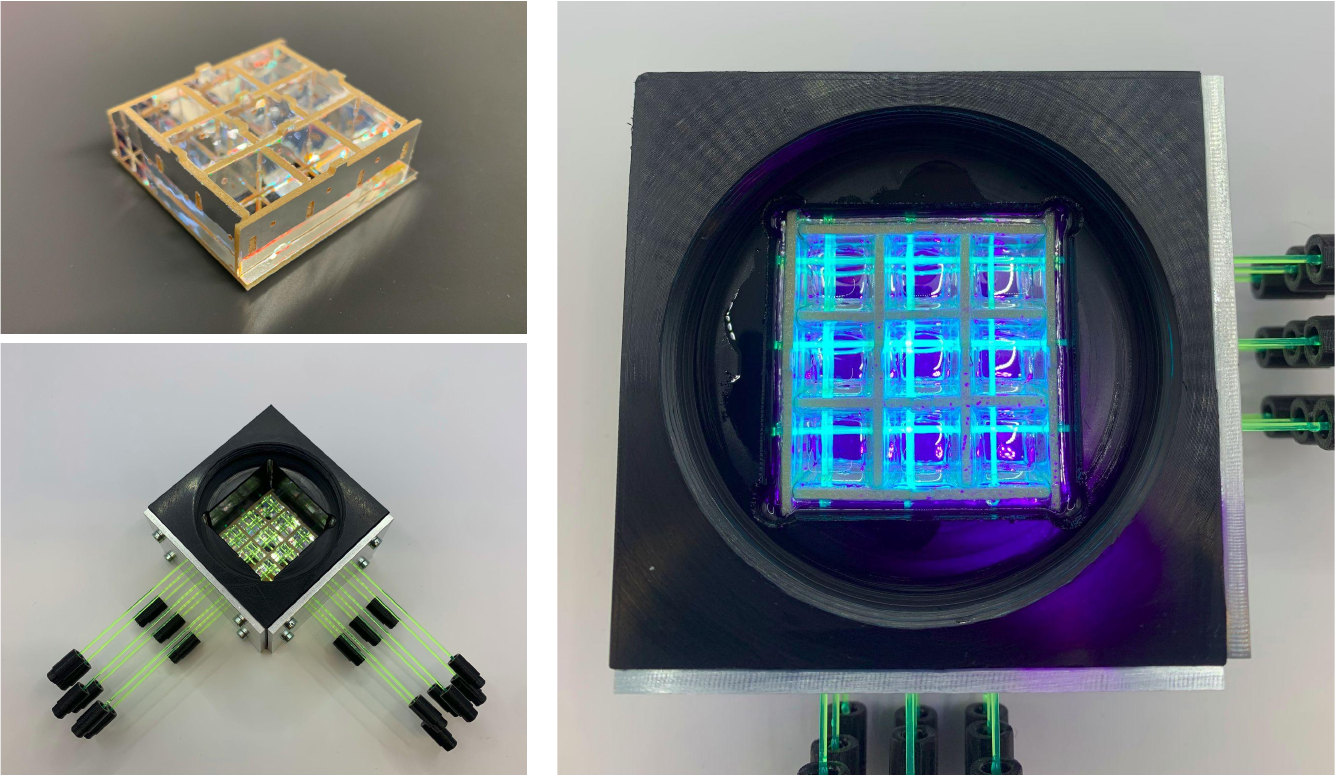}
    \caption{Top left: a single 3 $\times$ 3 matrix layer. Bottom left: the sealed dark box containing one layer of segmented WbLS and with six WLS fibers inserted. Right: the complete setup with all the three layers of segmented WbLS installed (top optical separator plane and top cover removed), exposed to UV light.
    }
	\label{fig:prototype}
\end{figure}

We built a prototype of $3 \times 3 \times 3$ optically-separated cubes of WbLS, made of three layers of $3 \times 3$ voxels, for a total of 27 cubes of WbLS. 
A second prototype with identical geometry was made, with the only difference that the WbLS was replaced with pure LAB+PPO+MSB LS. It was used as a reference to facilitate the evaluation of the optical performances of the WbLS prototype.
The assembly process involved sequentially placing each layer into the vessel, inserting the readout fibers, and filling the liquid, as shown in Fig.~\ref{fig:prototype} (bottom left). Each voxel was read out by two orthogonal wavelength-shifting (WLS) fibers, that provide two different projections of particle interactions. After all three layers were installed in the vessel, the top-most layer was covered with a single reflector piece to ensure complete optical isolation.
A total of 18 Kuraray Y11 WLS fibers were carefully polished at both ends to maximize signal transmission, and installed in the prototype. These fibers were later connected to Hamamatsu S13360-1350CS SiPMs~\cite{hamamatsu:mppc}, with a nominal photon detection efficiency (PDE) of 40\%. 
Custom 3D-printed connectors were used, and the coupling was ensured by a black foam pushing the SiPMs towards the fiber end. The analogue signal of the SiPMs was read out and digitized using a CAEN 
front-end board (FEB) DT5202~\cite{caen-FERS-board}.
This board provided fast data readout with a sampling rate of 2.5 ns and a nominal precision of 0.5~ns as well as trigger level discrimination.

\section{Characterization of the prototype optical performance}
\label{sec:measurements}

The detector concept described in Sec.~\ref{sec:conceptual-design} has been characterized, first, by measuring the reflectivity and the transmittance of the optical separator, then the light yield and the optical crosstalk of the $3 \times 3 \times 3$ cube prototype.

\subsection{Reflectivity and transmittance of the optical separators}
\label{sec:3m-reflector-measurement}

\begin{figure}[h!] 
	\begin{center}
		\includegraphics[width=0.7\textwidth]{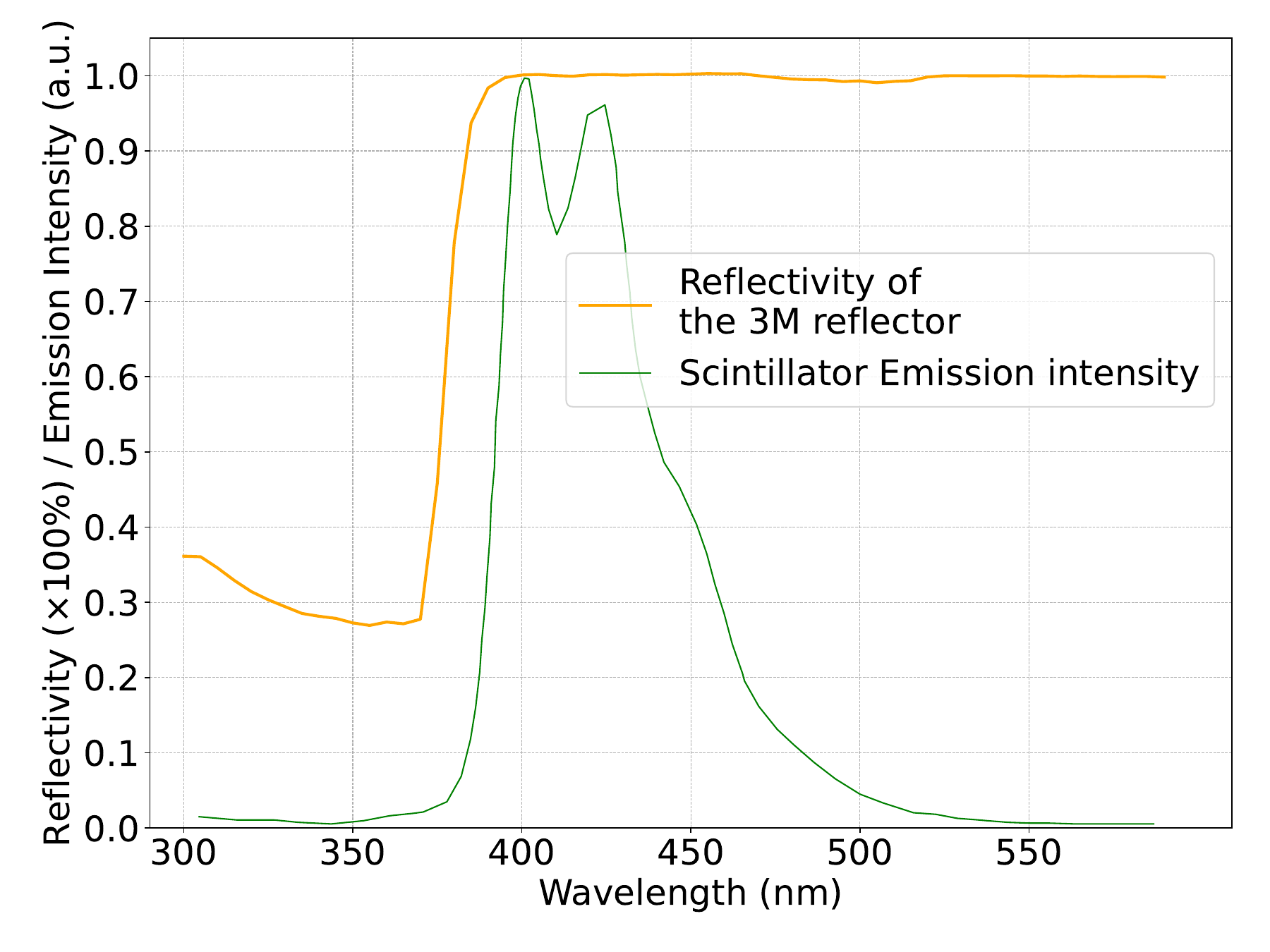}
		\caption{The measured reflectivity spectra of the customized 3M reflector, overlaid on top of the emission spectrum (in arbitrary units) of the utilized scintillator.
        }
		\label{fig:reflectivity}
	\end{center}
\end{figure}

The reflectivity and the transmittance of the customized reflector were measured using a Perkin Elmer Lambda 650 UV/VIS spectrometer~\cite{perkinelmer_lambda650}. Fig.~\ref{fig:reflectivity} shows the measured reflectivity spectrum, 
overlaid with the emission spectrum of the scintillator used in the prototype. 
The reflector in air exhibits high reflectivity ($>$99\%) across the entire scintillator emission spectrum, while no more than 0.3\% transmittance was measured. Furthermore, the customized reflector pieces were tested in 
the detector liquids
and no obvious performance degradation was observed. However, it should be noted that the spectrometer measurements were performed with the laser incident normal to the reflector surface.
Due to the features of the 3M specular reflector, that reflects light acting as a polarizer, higher transmittance may occur 
for large incident angles of the light ~\cite{3m_film_liquid}.
Moreover, being the reflector in contact with  materials of higher or similar refractive indices, such as WbLS and LS, the transmittance becomes higher. 
Since this measurement requires non-trivial modifications to the setup, we decided to leave it for future work and infer the reflectivity and transmittance directly from the prototype data (see Sec.~\ref{sec:opt-sim}).

\subsection{Scintillation light yield}
\label{sec:light-yield}

Both the WbLS and LS prototypes were independently tested with cosmic rays. An ``any-trigger'' scheme was employed during the test, in which all the 18 channels would start the data acquisition whenever any individual channel was triggered. This scheme ensured that the recorded charge distribution included not only 2D hits (i.e. channels with a certain number of photoelectrons counted by the SiPM) from primary particles but also hits from crosstalk and SiPM dark counts (the probability of having at least 1 dark count within the readout window of 100~ns is below 1\%). Primary particle 2D hits were used for prototype characterization, while crosstalk and dark counts were used to build the single-photon response profile for channel calibration. 
Each channel was calibrated as follows. The distribution of the signal measured in a single channel in units of analogue-to-digital converter (ADC) showed a clear multi-peak structure below 600 ADC, each one corresponding to a different number of photoelectrons (PE), which represent the primary electrons originated via photoelectric effect by the visible photons impinging on the SiPM active area.
The SiPM gain is computed as the distance between two adjacent peaks in units of ADC per p.e. with a multi-Gaussian distribution fit. Thus, the position of the first three peaks was obtained and fitted with a linear function and both the gain and the pedestal were extracted.

Cosmic rays are MIPs and deposit energy in LS and WbLS proportional to their path length in the active volume, around 
2 MeV/cm. 
Therefore, single, straight-through particle tracks were retained for their consistent path length of about 1 cm in each cube voxel. The following selection procedure was applied on an event-by-event basis: in each layer (perpendicular to the Z-axis, the primary cosmic axis), a unique 3D hit candidate was reconstructed by matching the highest-charge 2D hits from the two detector projections. The matched 2D hits must surpass the charge threshold (5 p.e. for LS and 1.5 p.e. for WbLS), optimized to suppress crosstalk and SiPM dark counts, while retaining true track signals. If no 3D hit could be reconstructed for a given layer, the event was rejected. This selection ensured a single through-going track for each selected event. A second selection step was applied to retain only low-angle tracks by requiring that the reconstructed 3D hits in each layer shared identical X–Y coordinates. This ensured a consistent path length through each voxel and, consequently, a uniform detector response.
%
The light yield per WLS fiber was evaluated for single-channel signals, per voxel (average between the two corresponding readout channels), and per track (average of six channels along each selected track).
\begin{figure*}[tp!]
	\centering 
    \begin{subfigure}[t]{0.45\textwidth}
        \centering
        \includegraphics[width=\textwidth, angle=0]{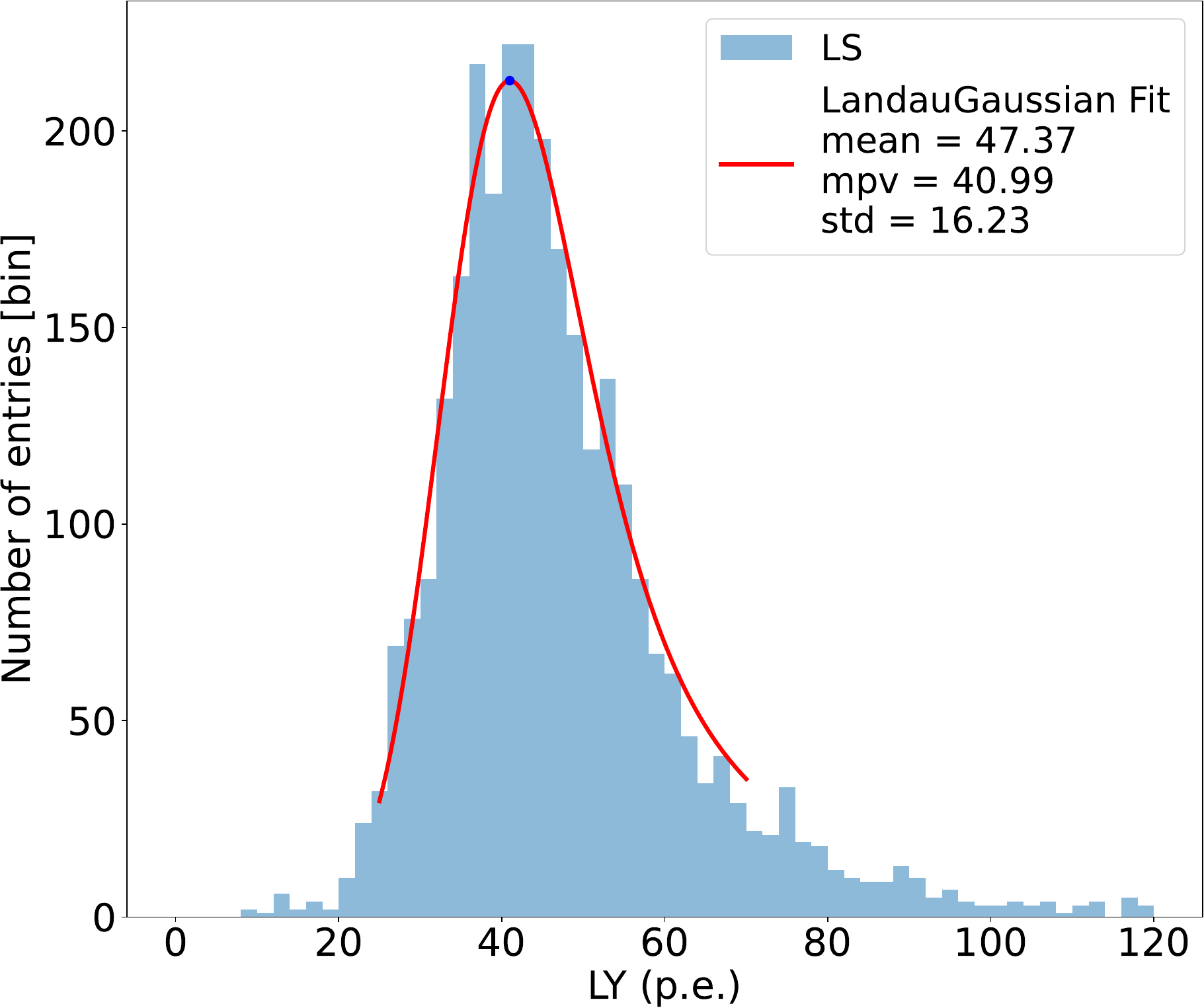}
    \end{subfigure}
    \begin{subfigure}[t]{0.45\textwidth}
        \centering
        \includegraphics[width=\textwidth, angle=0]{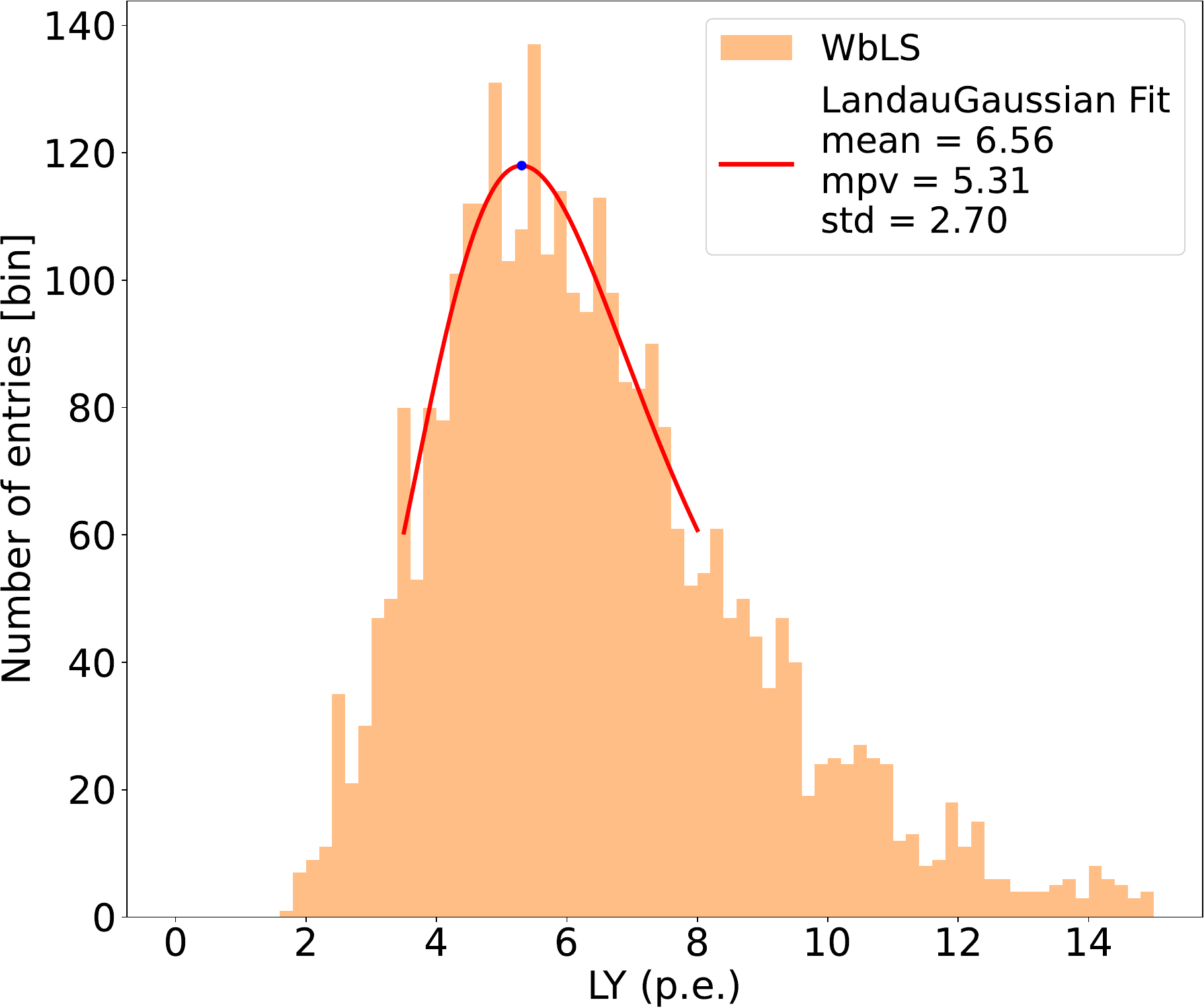}
    \end{subfigure} 
    \begin{subfigure}[t]{0.45\textwidth}
        \centering
        \includegraphics[width=\textwidth, angle=0]{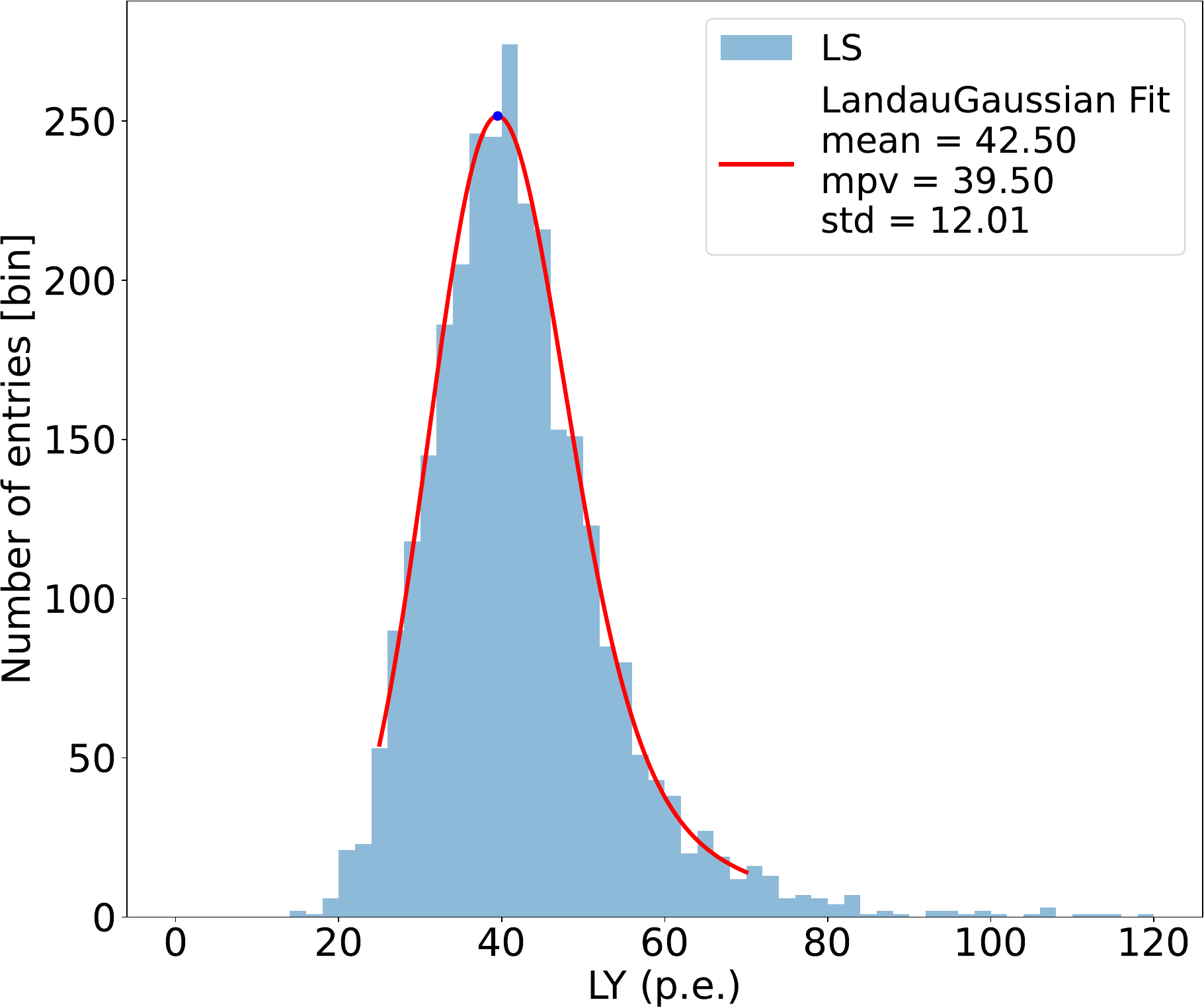}
    \end{subfigure}
    \begin{subfigure}[t]{0.45\textwidth}
        \centering
        \includegraphics[width=\textwidth, angle=0]{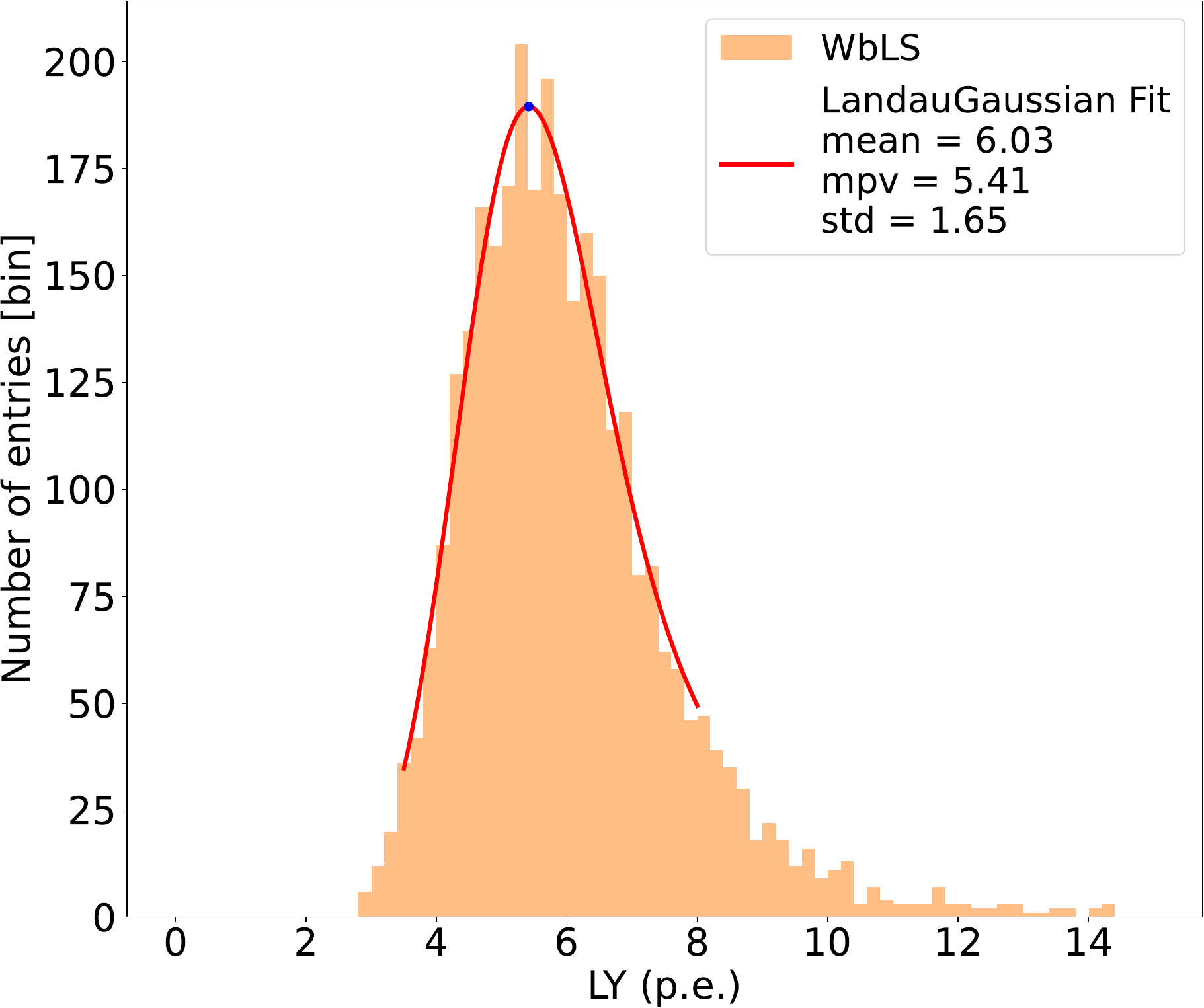}
    \end{subfigure} 
    \caption{
    Distributions of the light yield in units of p.e. for LS (left) and WbLS (right).
    The top histograms are obtained by averaging the light yield of two channels of a same cube, while the bottom histograms are obtained by averaging the light yield of all the six channels fired by the selected cosmic ray track.
    The standard deviation (std) and the mean value of the histogram are shown.
    The distributions were fitted with a Landau-Gaussian function to find the peak position and infer the most-probable value (mpv). 
    }
	\label{fig:ly}
\end{figure*} 

Fig.~\ref{fig:ly} shows the averaged channel-wise light yield distributions for WbLS and LS. Each distribution was fitted with a Landau-Gaussian function. 
The most-probable value (mpv) of the light yield was estimated from the average of six channels and was found to be 5.4 p.e./channel/MIP for WbLS and 39.1 p.e./channel/MIP for LS 
with an average light yield of 6.0 p.e./channel/MIP for WbLS and 42.5 p.e./channel/MIP for LS. 
This ratio is consistent with expectations, given that the tested WbLS contains approximately 10\% scintillator by mass. 
The small deviation might come from the difference in the refractive indices of LS and WbLS, where more light is trapped in the WLS fiber due to the larger refractive index contrast between water and the fiber material
as well as from the different fractional contributions of the Cherenkov light (a detailed discussion can be found in Sec.~\ref{sec:opt-sim-results}).
Additional contributions might arise from possible non-uniformities in the precision of the SiPM to WLS fiber coupling. 

Overall, the measured WbLS light yield is sufficient for charged particle tracking: MIP signals are well above the electronic pedestal, and protons are expected to produce up to five times more light near the Bragg peak. However, further studies are required to fully assess the tracking performance of a large-scale WbLS-based detector.
The fractional width (standard deviation / mpv) of the light yield averaged over the six channels of the track is 30\% for both WbLS and LS. It increases to 40\% and 51\%, respectively for LS and WbLS, when only the two channels from a single voxel are taken into account in the computation. 
From preliminary calculations we see that, while the Poisson fluctuation of the light yield of WbLS is larger than that of LS when averaging over two channels, this contribution is suppressed when averaging six channels, resulting in the same fractional width.
Overall, this seems to highlight a 
limited 
deterioration of the particle identification in WbLS compared to LS that will need to be verified with physics simulation with the proper setup of a GeV neutrino experiment.
The light yield of LS is comparable to those observed in the state-of-the-art plastic scintillator detectors~\cite{Blondel:2020sfgd}.

\subsection{Cube-to-cube optical crosstalk}
\label{sec:crosstalk}

The crosstalk is defined as the light escaped through one of the six faces of the voxel activated by a charged particle and captured by a WLS fiber of an adjacent voxel and counted by the corresponding SiPM. Generally, the crosstalk rate can be approximated as the ratio between the light yield measured by one channel of the activated voxel and that measured by the parallel readout channel in an adjacent voxel. A too large crosstalk rate would worsen the spatial resolution of the detector and weaken its capability as a particle tracker. Thus, it is important to constrain the crosstalk to a few percent level.

\begin{figure*}[t!]
	\centering 
    \begin{subfigure}[t]{0.45\textwidth}
        \centering
        \includegraphics[width=\textwidth, angle=0]{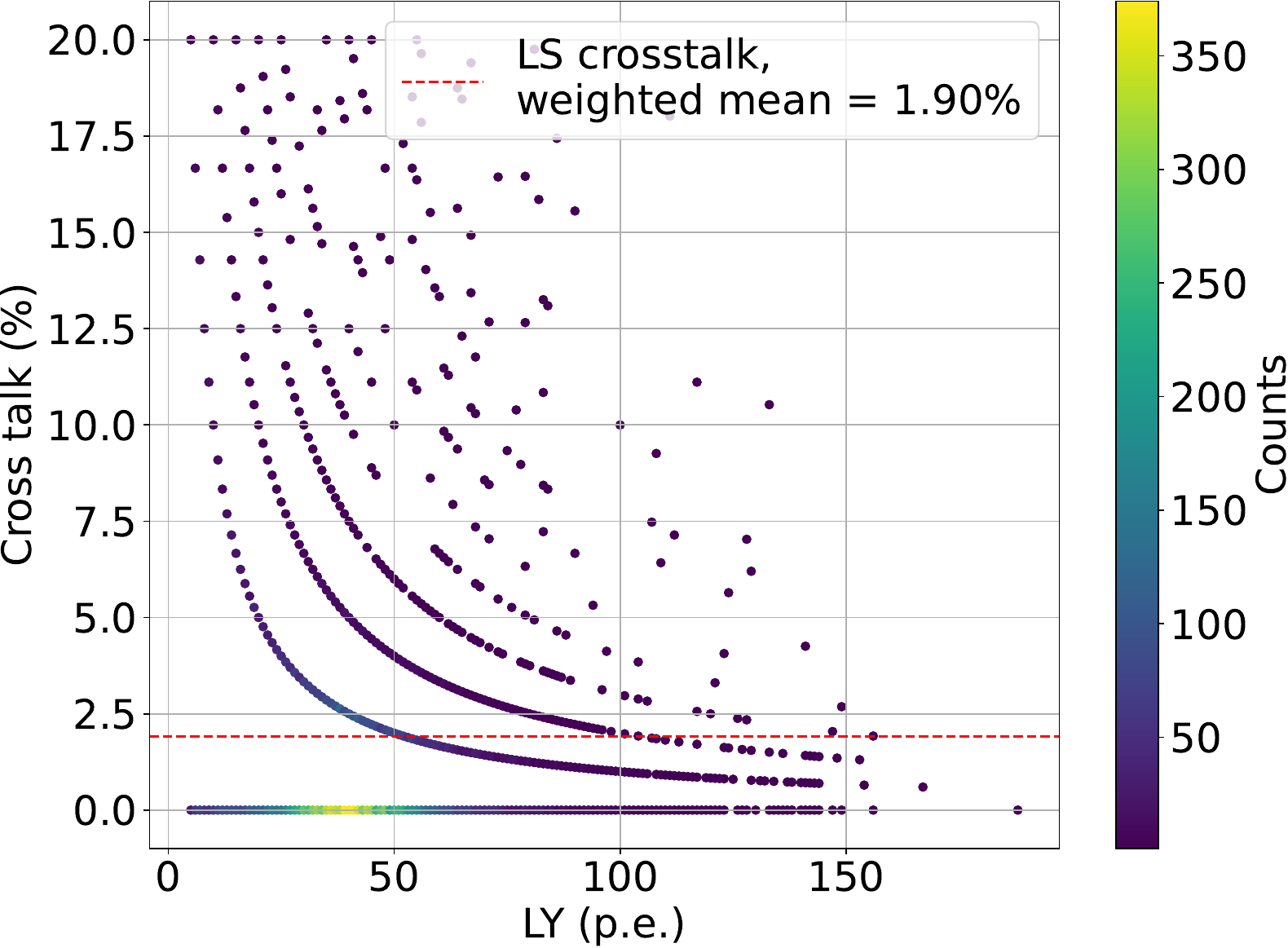}
    \end{subfigure}
    \begin{subfigure}[t]{0.45\textwidth}
        \centering
        \includegraphics[width=\textwidth, angle=0]{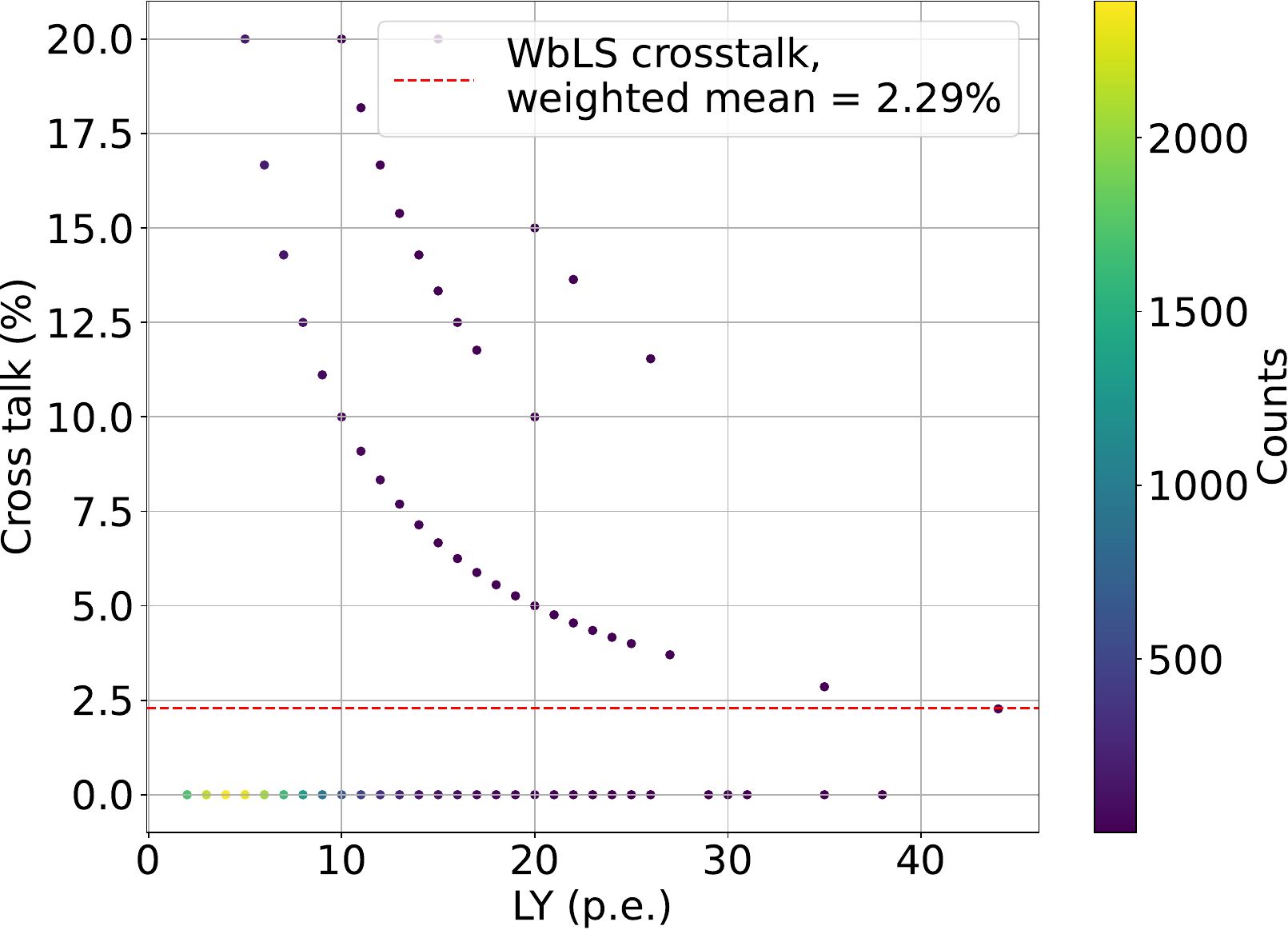}
    \end{subfigure}
    \caption{The distributions of the measured crosstalk rate versus the light yield of LS (left) and WbLS (right). All the light yield data were rounded to integer p.e.. The number of events measuring the same light yield and crosstalk was reflected by the color scale. The red dashed line illustrates the average crosstalk rate estimates.
    }
	\label{fig:crosstalk}
\end{figure*}

Due to the optical isolation provided by the customized reflectors, the setups were expected to exhibit minimal cube-to-cube crosstalk.
For WbLS, where the light yield in the primary cube is relatively low, most crosstalk in adjacent cubes is typically expected to fall below the detection threshold of 1 p.e..
To visualize this effect, 2D histograms (see Fig.~\ref{fig:crosstalk}) were used to show the crosstalk distribution as a function of the main channel light yield. The crosstalk (vertical axis) is expressed as a percentage, while the main channel light yield (horizontal axis) is reported in units of p.e.. The left plot shows the distribution of the LS, while the results of the WbLS are shown in the right plot.

The measured light yields were rounded to integer p.e., resulting in discrete patterns in the distributions. Each data point represents an event where the corresponding light yield and crosstalk were measured, and the number of events measuring the same results is reflected by the color. As expected, in both distributions, most events are concentrated along the horizontal line, corresponding to zero measured crosstalk. The red dashed lines represent the estimated crosstalk rate by averaging over all the events, using the following formula:
\begin{equation}
\text{Crosstalk} = \sum_{i=1}^{i=N_{evt}}\text{LY}_{i}(\text{neighbor}) / \sum_{i=1}^{i=N_{evt}}\text{LY}_{i}(\text{main}),
\end{equation}
where $\text{LY}_{i}(\text{main})$ is the light yield measured by a channel of the voxel hit by the particle, $\text{LY}_{i}(\text{neighbor})$ is the light yield measured by the corresponding parallel channel in an adjacent voxel, i.e., the crosstalk signal. The sum runs over all the selected events ($N_{evt}$). Overall, a crosstalk rate of 1.90\% was obtained for the LS and 2.29\% for the WbLS. 
Due to the lower electronics threshold 
in the WbLS prototype, pedestal fluctuations and SiPM dark counts were more likely to be misidentified as crosstalk signals. This effect was less pronounced in LS, given the higher threshold, where true non-zero crosstalk hits were more frequent. As a result, this method likely overestimated the true crosstalk rate in the WbLS configuration.

In summary, both setups exhibited low
levels of cube-to-cube crosstalk, around 2\%,
that is lower than the similar 3D granularity detectors available in the literature~\cite{Blondel:2020sfgd,Mineev:2018ekk}. 



\section{Optical model of the water-based liquid scintillator prototype}
\label{sec:opt-sim}

\begin{figure}[h]
	\centering 
    \includegraphics[width=0.5\textwidth, angle=0]{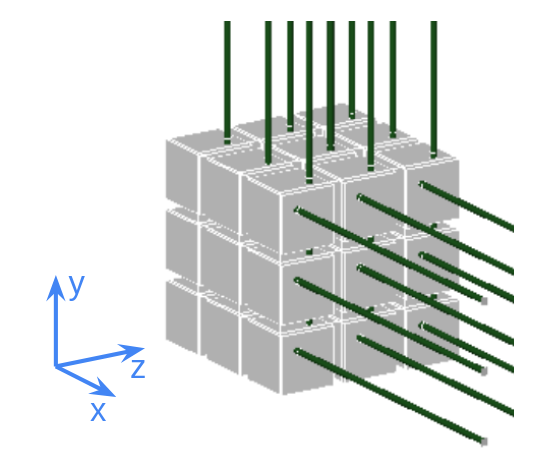}
    \caption{The prototype geometry simulated in Geant4.
    The simulated cosmic rays travel from negative to positive Z.
    }
	\label{fig:MC_Setup}
\end{figure}

A detailed optical simulation has been developed in Geant4
~\cite{Allison_2006_Geant4_developments_and,Agostinelli_2003_Geant4_a_simulation_toolkit,Allison_2016_Recent_developments_in}
to estimate the effective optical parameters of the prototype, such as the reflectivity and the transmittance of the optical separator. 
An overview of the simulated geometry can be seen in Fig.~\ref{fig:MC_Setup}. 
The geometry includes 27 cube-shaped detector voxels. Each voxel comprises an outer reflector matrix, an inner active volume of 1 $\times$ 1 $\times$ 1 cm$^3$ filled with either WbLS or LS (same composition of the prototype), and two 9 cm long orthogonal WLS fibers in direct contact with the scintillator volume. These fibers are read out at one end by a SiPM volume with the nominal PDE spectrum \cite{hamamatsu:mppc} implemented. 
The model parameters and the corresponding range of allowed values are: 
reflectivity $R$ (77 - 90\%), transmittance $T$ (0.5 - 7\%), and the primary light outputs of LS (8000 - 11,500 photons/MeV) and WbLS (800 - 1150 photons/MeV), and scintillator attenuation length (20 m).
The effective absorption length of the optical separator (3M reflector plus divinycell) is defined as $1-T-R$. 
The intervals of the parameters were defined based on a first coarse scanning for a preliminary evaluation of the goodness of fit (see method in Sec.~\ref{sec:opt-sim-results}).

To emulate the behavior of cosmic-ray events,
muons were simulated with kinetic energies uniformly distributed between 300 and 650 MeV, emitted within a narrow forward cone
perpendicular to the Z-axis, starting 6 mm upstream of the detector volume
and entering from the external surface (1 $\times$ 1 cm$^2$) of the central cube. 

The same selection of cosmic rays as for the prototype data was implemented.
Straight, through-going tracks were selected and reconstructed using the methods described in Sec.~\ref{sec:measurements}. An additional fiducial volume cut was applied, requiring tracks to be fully contained within the three central cubes along the particle direction. Furthermore, the cut requires that the 
light yield (number of p.e.)
in neighboring channels remain below 50\% of that in the corresponding on-track channels, thereby suppressing events with spurious optical crosstalk caused by off-axis track segments. Events passing all selection criteria were retained for the following analysis.

\subsection{Results of the optical simulation}
\label{sec:opt-sim-results}

First, we focused on extracting the optical transmittance ($T$) and reflectivity ($R$) of the customized reflector as well as the scintillator intrinsic yield, that we refer to as light output. These parameters were inferred by comparing the measured and the simulated distributions 
of the light yield (see Sec.~\ref{sec:light-yield}) and the optical crosstalk (see Sec.~\ref{sec:crosstalk}).
A 2-dimensional template analysis fit was implemented by comparing the detected light yield as a function of the optical crosstalk per channel, both in units of p.e., with the one simulated for different hypotheses of $R$, $T$ and light output.

\begin{figure}[h!]
    \centering
    \begin{subfigure}[t]{0.45\linewidth}
        \centering
        \includegraphics[width=\textwidth]{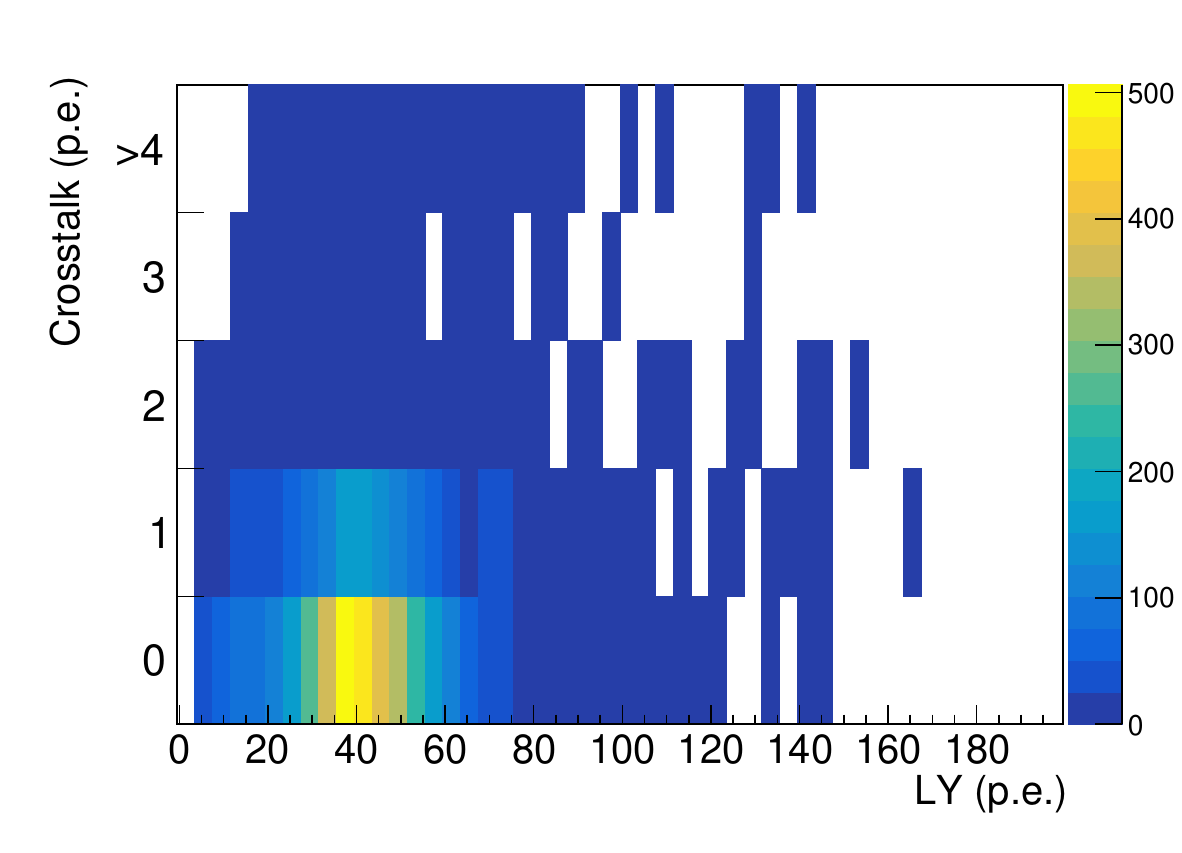}
        \caption{LS measurement data}
    \end{subfigure}
    \hfill
    \begin{subfigure}[t]{0.45\linewidth}
        \centering
        \includegraphics[width=\linewidth]{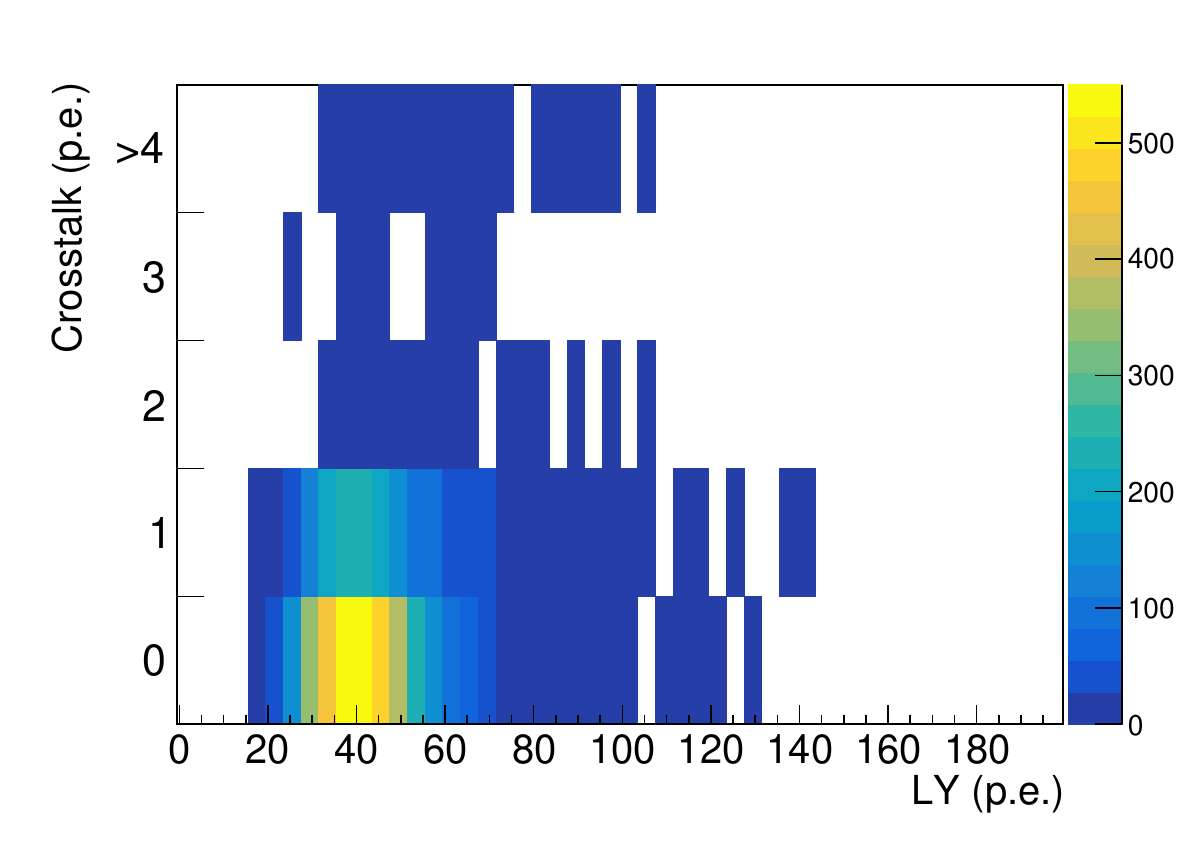}
        \caption{LS MC data (R = 86\%, T = 1.5\%)}
    \end{subfigure}
    \caption{The 2D distributions of the light yield and optical crosstalk in units of p.e. are shown.
    The LS prototype data (left) and the MC best fit with a primary light output of 9'500 photons/MeV (right) histograms are shown. Events with a crosstalk of 4 p.e. or more were merged into a single bin to increase the statistics.
    }
    \label{fig:2d_data_LS}
\end{figure}

\begin{figure}[h!]
    \begin{subfigure}[t]{0.45\textwidth}
        \centering
        \includegraphics[width=\textwidth]{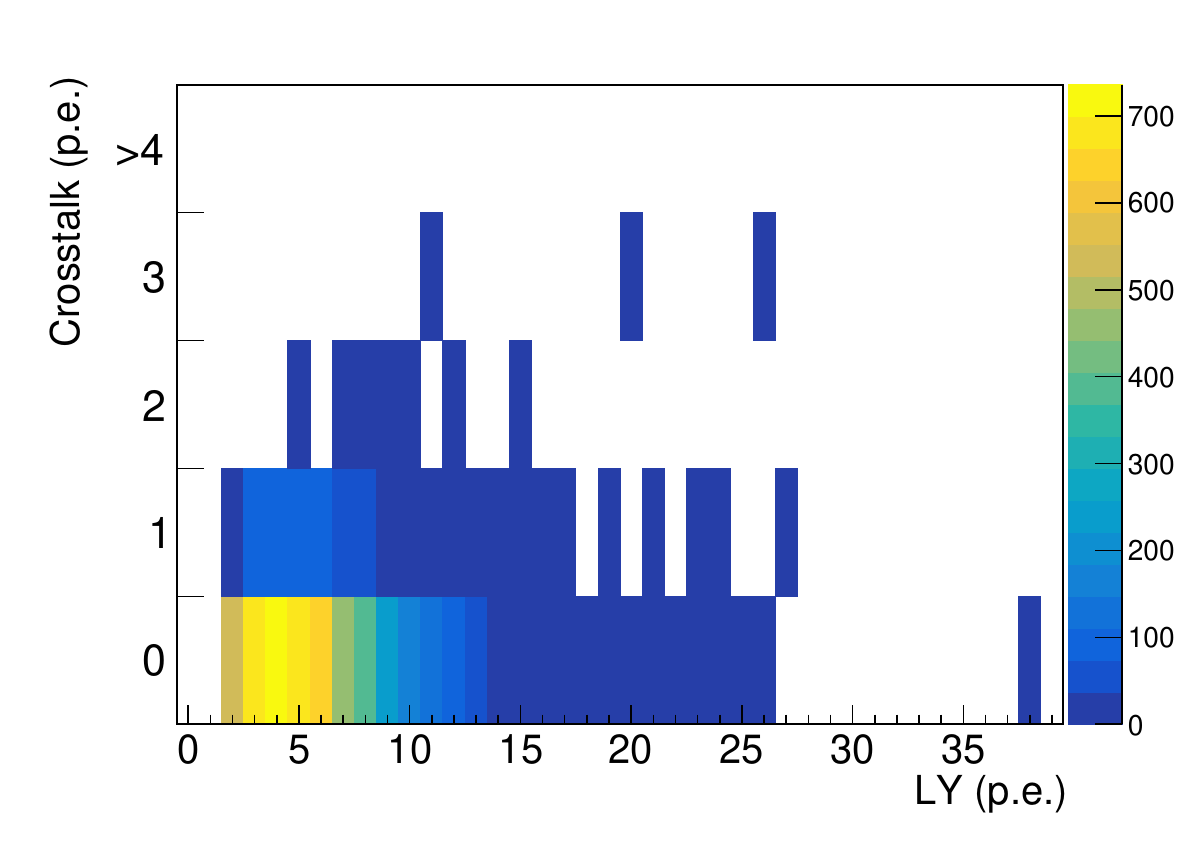}
        \caption{WbLS measurement data}
    \end{subfigure}
    \hfill
    \begin{subfigure}[t]{0.45\textwidth}
        \centering
        \includegraphics[width=\textwidth]{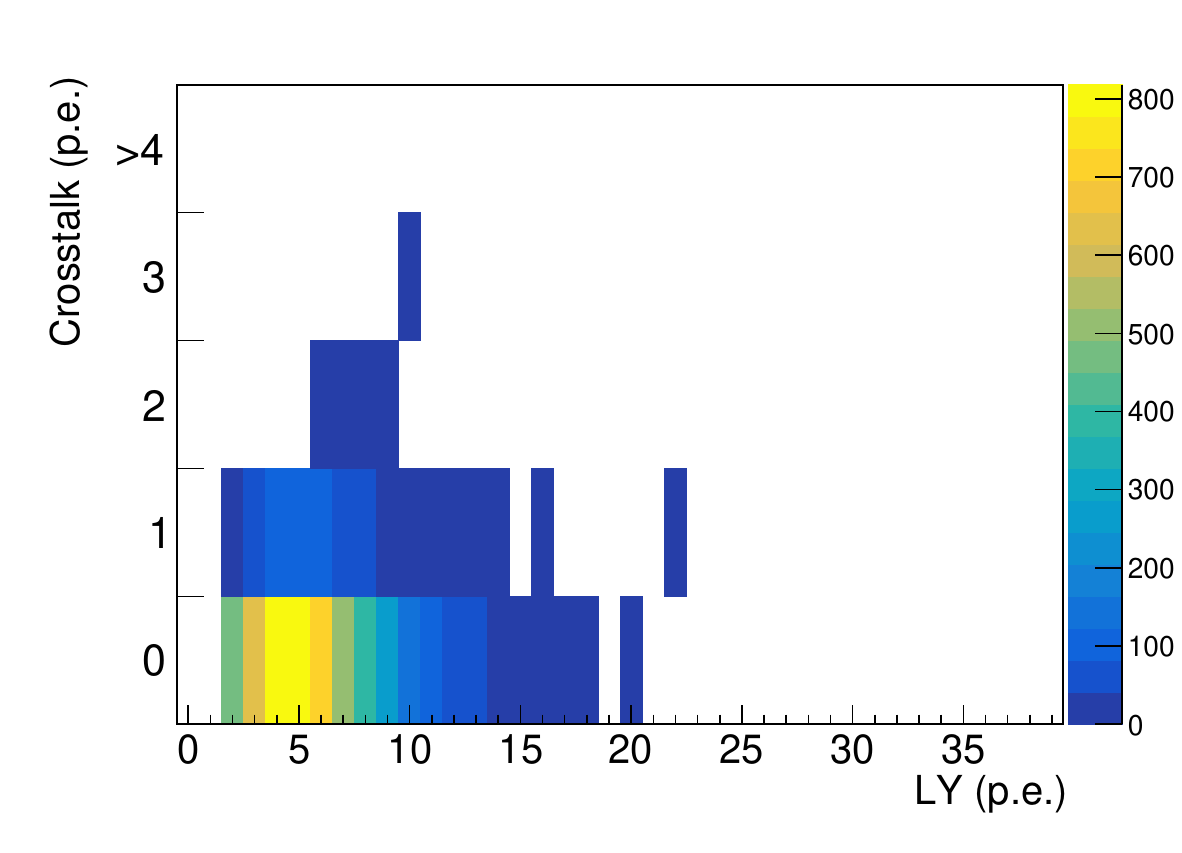}
        \caption{WbLS MC data (R = 85\%, T = 2.5\%)}
    \end{subfigure}
\caption{The 2D distributions of the light yield and optical crosstalk in unit of p.e. are shown.
The WbLS prototype data (left) and the MC best fit with a primary light output of 950 photons/MeV (right) histograms are shown. Events with a crosstalk of 4 p.e. or more were merged into a single bin to increase the statistics.}
\label{fig:2d_data_WbLS}
\end{figure}

The following fitting scheme was adopted:
a 2D histogram of light yield versus crosstalk is obtained from the optical simulation for a given hypothesis of $R$, $T$ and light output; 
the goodness of fit for the given set parameter values was quantified by computing the chi-square defined as
\begin{equation}
\chi^2=2\sum_{i,j}\frac{(H_{data}^{i,j} - H_{MC}^{i,j})^2}{H_{data}^{i,j}+H_{MC}^{i,j}},
\label{WbLS_chi2}
\end{equation}
where $H_{data}(i,j)$ and $H_{MC}(i,j)$ are the data and the simulation bin values for the i,j-th bin of the 2D histogram.
The procedure was iteratively performed for different values of the reflectivity, transmittance and light output parameters.
The set of parameter values corresponding to the minimum $\chi^2$ was identified as the best fit, corresponding to the best agreement between the simulated model and the experimental data. 
A representative example is shown in  Fig.~\ref{fig:2d_data_LS} and~\ref{fig:2d_data_WbLS}.

\begin{figure*}[t!]
	\centering
    \begin{subfigure}[t]{0.45\textwidth}
        \centering
        \includegraphics[width=\textwidth, angle=0]{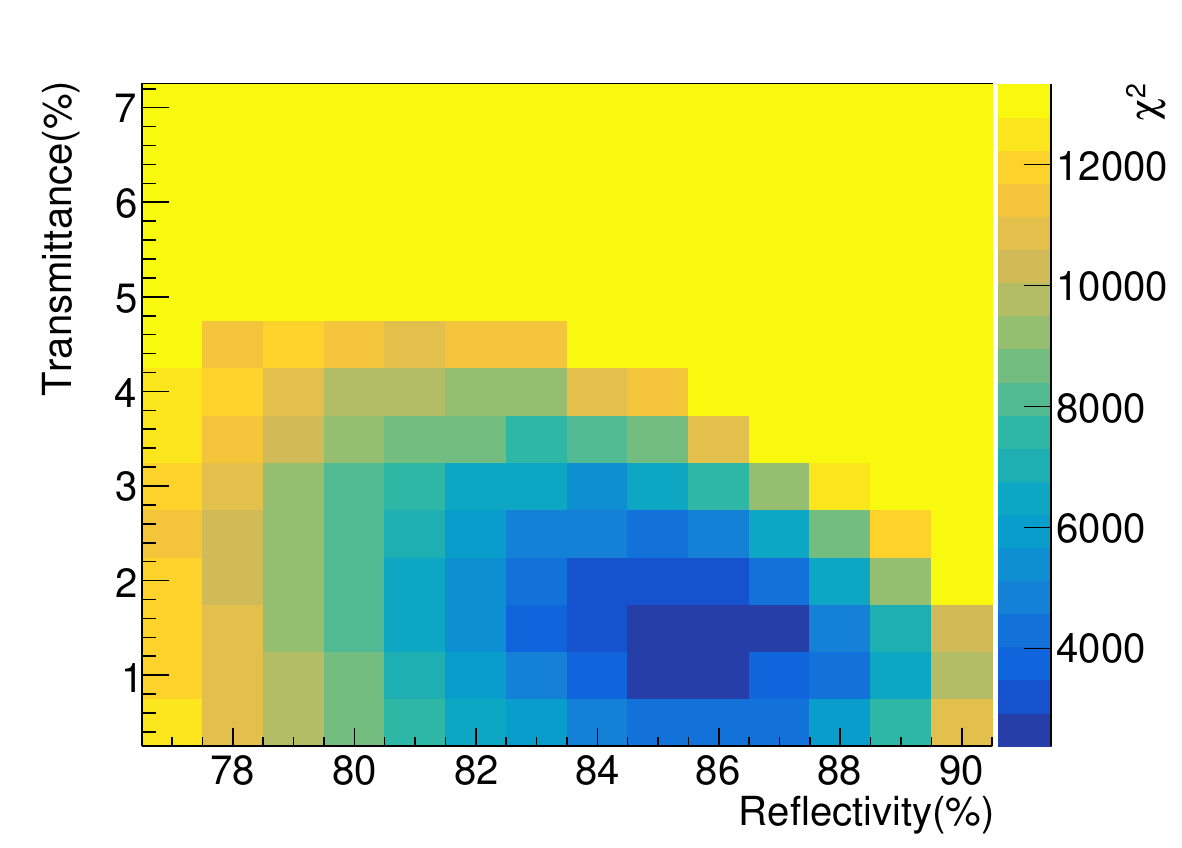}
        \caption{The $\chi^2$ distribution of the LS fit.}
    \end{subfigure}
    \begin{subfigure}[t]{0.45\textwidth}
        \centering
        \includegraphics[width=\textwidth, angle=0]{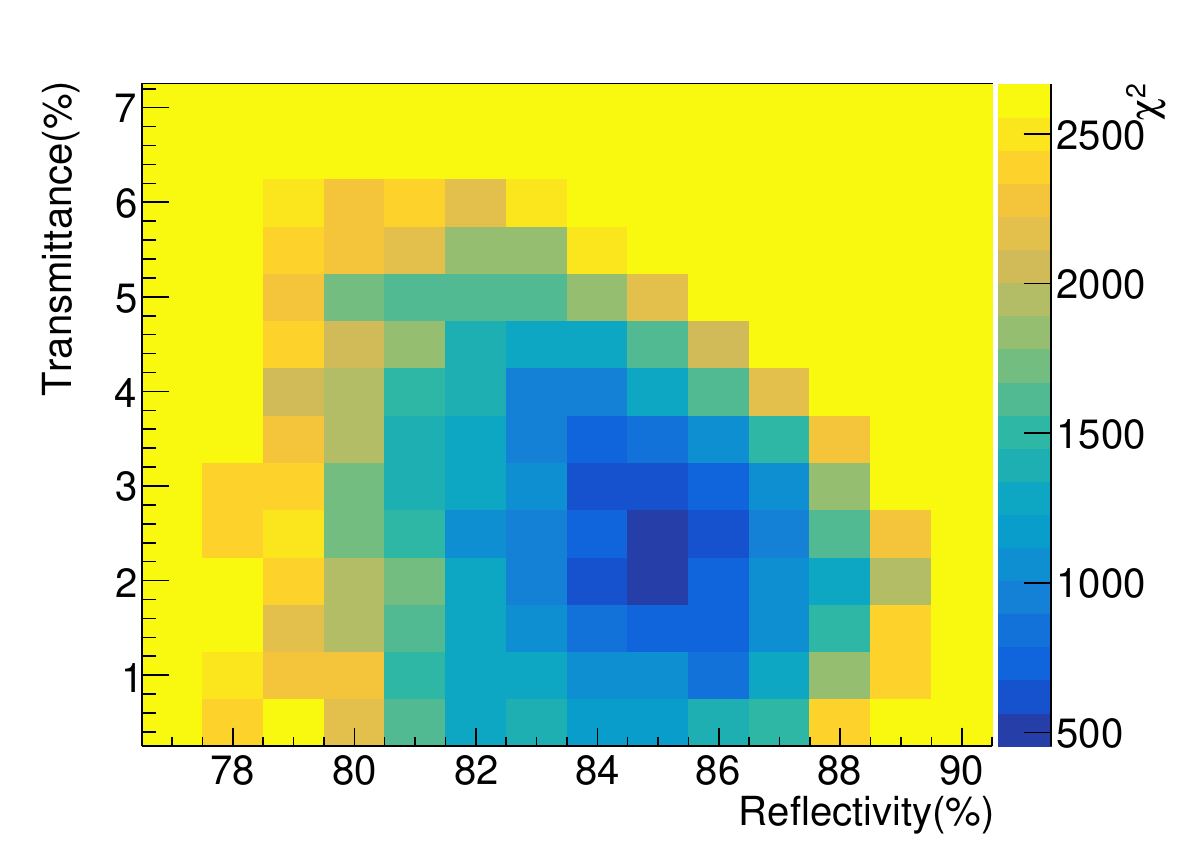}
        \caption{The $\chi^2$ distribution of the WbLS fit.}
    \end{subfigure}
    \caption{The $\chi^2$ distributions corresponding to the best fit parameters for the LS (left) and WbLS (right) prototypes. The light output was fixed, respectively, to the nominal values of 9500 ph/MeV and 950 ph/MeV.} 
	\label{fig:chi2_map}
\end{figure*}

The fits of the WbLS and LS prototype data were performed independently, as the effective optical properties of the reflective film may differ depending on the refractive index of the liquid as well as the different light incident angles. 
The scintillation light output was constrained by measurements available in literature for similar LS formulations~\cite{SNO+, DayaBay} and by the known 10\% concentration by mass of LS in the WbLS sample.
As an example, Fig.~\ref{fig:chi2_map} shows the resulting $\chi^2$ maps for LS and WbLS, where the light outputs were fixed to 9500 photons/MeV and 950 photons/MeV, respectively.
Degeneracies can arise between the different parameters, for example in the anti-correlation between the reflectivity of the separators and the intrinsic light output of the WbLS.   

For LS, the best fit at a primary light output of 9500 photons/MeV corresponded to $R = 86\%$ and $T = 1.5\%$, with a $\chi^2$ of 2383.6. For WbLS, the best fit at 950 photons/MeV was obtained with a reflectivity of 85\% and a transmittance of 2.5\%, and corresponded to a $\chi^2$ of 454.6.
This fitting procedure was repeated across the full light output range, resulting in an optimal $(R, T)$ pair for each assumed value. Overall, the fitting results exhibited a high degree of degeneracy, with only minor variations observed in the $\chi^2$ values of the corresponding optimal parameter sets. The extracted transmittance parameter $T$ remained tightly constrained by the measured optical crosstalk, varying between 1–2\% across the light yield range. In contrast, the effective reflectivity $R$ showed a broader allowed range, consistently falling between 81\% and 89\%, with a relatively good agreement observed between the LS and WbLS configurations. 


The production of the Cherenkov light as well as its absorption in the WLS fibers were simulated in Geant4.
Its contribution to the total light yield was found to be about 0.5 p.e. per channel, thus smaller than 10\% of that of scintillation light in WbLS.
On the other hand, it is worth noting that a certain fraction of the produced Cherenkov light can be absorbed by the doping loaded in the scintillator and emitted with the same wavelength as the primary scintillation light.
The simulation of this component is non-trivial as it requires an accurate knowledge of the doping absorption length independently for WbLS and LS. Thus, it was accounted for as a contribution to the total scintillation light and tuned on the cosmic ray data.
The simulation shows that 
about 190 
photons 
are generated in a 1 cm voxel by a 1 GeV muon 
in the wavelength range covered by PPO and MSB. Conservatively assuming a 100\% absorption/emission efficiency and a scintillation light output of 950 photons per MeV in WbLS, we can set an upper limit for this component 
to about 
10\%
of the total scintillation light
($\sim 2$ MeV loss per centimeter).

In summary, the optimized set of parameters indicates that the effective reflectivity of the customized reflector in liquid was significantly lower than the 99\% measured in air. The high reflectivity of the 3M film relies on constructive interference, which is optimized for a high refractive index contrast between the film material and air. However, this effect can be significantly reduced when the film is in contact with high-refractive-index materials such as LS or WbLS. Moreover, the maximum of the reflectivity is for light normal to the reflecting surface, while it could degrade with higher incidental angles, as noted in Sec.~\ref{sec:3m-reflector-measurement}.
Nonetheless, this study provided a comprehensive understanding of segmented liquid scintillator prototypes, and the characterization of the customized reflector offers valuable insight for the future optimization of optical isolation designs.

\section{Conclusions}
\label{sec:}

Precision detection of neutrino-nucleus interactions in water with the complete detection of the final state, including leptons and hadrons, has always been a challenge given that water does not scintillate. This has been a limitation in the constraint of the neutrino-nucleus interaction models.
The invention of water-based liquid scintillator can be a game changer for the near detectors of the future long-baseline neutrino oscillation experiments with water-Cherenkov far detectors, such as Hyper-Kamiokande, the proposed ESS$\nu$B experiment, or the THEIA far detector 
proposed for DUNE. 

We propose a novel design consisting of a 3D highly-segmented water-based liquid scintillator with the aim of precision detection of neutrino-nucleus interactions in water. The water-based liquid scintillator is encapsulated within a highly-segmented rigid but very light optically-isolating structure.
Each voxel is optically isolated by a customized reflector made of a low-density rigid support sandwiched by highly-reflective thin mirror layers and read out by orthogonal wavelength-shifting fibers.
Such configuration is also suitable for pure liquid scintillator. 

The prototype was tested with cosmic ray data and compared to the Geant4-based optical simulation to gain insights about the optical model and best-fit effective parameters, and to get guidance towards a further optimization of the detector in the future.
An average light yield of 6.0 p.e./channel/MIP was measured with the WbLS prototype, to be compared with 42.5 p.e./channel/MIP for the LS one. The measured light yield is consistent with the 
nominal
expectations, underscoring the stability of the production process. 
Despite the lower light yield of WbLS, the obtained results hint at a not dramatic worsening of the PID capability by dE/dx of the 3D segmented WbLS. Dedicated studies will be performed in the future.
It is found that the light yield of WbLS is sufficient for physics applications, such as detecting MIP-like particles generated in the neutrino interactions, including muons, electrons and pions. The light yield will increase by approximately a factor of 5 for stopping protons showing the Bragg-peak feature. 
Furthermore, the cube-to-cube optical crosstalk rates of 1.90\% for LS and 2.29\% for WbLS highlight the effectiveness of the reflective isolation design, ensuring the spatial resolution envisaged by design. 
This is key to distinguishing, counting and reconstructing the range of multiple low-momentum protons stopping near the neutrino interaction vertex.  
This 3D-granularity detector configuration is certainly suitable also for a pure LS detector, potentially cheaper than plastic scintillator.


This work highlights the potential of 3D-segmented WbLS and LS as a versatile medium for neutrino detectors. 
In future prototypes, the vertical WLS fiber (not included in this work) can be easily added to provide the third projection of the particle interaction.
It will also aim to increase the number of p.e..
Several promising options will be tested: 
SiPMs with higher photodetection efficiency, such as those with pixels as large as 75 $\mu$m \cite{hamamatsu:mppc}, aiming for an additional relative increase of about 20\% compared to this work;
higher SiPM bias voltage, whose independent prototype tests showed an increase of the light yield by 25\%;
double the number of WLS fibers per projection (only 1\% reduction of water per additional fiber) coupled with larger-surface SiPMs (e.g. $3 \times 3$ mm$^2$), that from simulations is expected to increase the light yield by about 60\%;
highly reflective mirror paint at the WLS fiber end.
Overall, a single-channel light yield up to 10 p.e. or more per SiPM can be expected. 
Moreover, the purging of the WbLS (not done in this work), the optimization of the optical reflector film as well as of the WbLS recipe (while maintaining the same fraction of water) is expected to further improve the detector light yield.
Finally, the 81\% of water by mass in the active volume and the 0.73 g/cm$^3$ can be increased to values above, respectively, 85\% and 0.87 g/cm$^3$ by using thinner foam (e.g. 0.5 mm) and reflector (e.g. 0.05 mm).

\bibliographystyle{apsrev4-2}
\bibliography{Reference}

\section{Acknowledgements}

We thank the CERN scintillator group, in particular Thomas Schneider, for giving us access to their spectrometer for the reflectivity and transmittance measurements.
We thank the members of the Hyper-Kamiokande ND280++ working group for the useful discussions of the results.
Part of this work was supported by the grants PCEFP2\_203261 and 200021L-231581 of the Swiss National Science Foundation.

\subsection{Contributions}

DS and TK conceived the detector concept.
BL, DS, TW, MY designed the specific detector configuration and prototyped it.
DS supervised the R\&D work, measurements and data analyses.
MY developed and provided the pure liquid and water-based liquid scintillators and supervised the R\&D. 
BL built the prototype, set up and performed all the measurements reported, both with the detector prototype and reflectivity and transmittance once, performed the data analysis, developed and ran the optical simulation.
TW and JW built the prototype.
UK set up and performed the reflectivity and transmittance measurements.
DB and MF helped to run the measurements.
DB performed the data analysis.
DB obtained the results of the optical simulation.
Also AR supervised the data analysis and simulation studies. 
CG and TK provided useful contribution on the understanding of the results.



\subsection{Competing interests}
The authors declare no competing interests.

\end{document}